\documentclass[preprint,authoryear,12pt]{elsarticle}

\usepackage{lineno,hyperref}
\modulolinenumbers[5]

\begin{document}
\begin{frontmatter}

\title{Effect of dark matter halo on global spiral modes in a collisionless galactic disk}
\author[mymainaddress]{Soumavo Ghosh\corref{mycorrespondingauthor}}
\cortext[mycorrespondingauthor]{Corresponding author}
\ead{soumavo@physics.iisc.ernet.in}

\author[mymainaddress,mysecondaryaddress]{Tarun Deep Saini}
\ead{tarun@physics.iisc.ernet.in}

\author[mymainaddress,mysecondaryaddress]{Chanda J Jog}
\ead{cjjog@physics.iisc.ernet.in}

\address[mymainaddress]{Department of Physics, Indian Institute of Science,Bangalore 560012, India}

 \begin{abstract}
 Low surface brightness (LSB) galaxies are dominated by dark matter halo from the innermost radii; hence they are ideal candidates to investigate the influence of dark matter on different dynamical aspects of spiral galaxies. Here, we study the effect of dark matter halo on grand-design, $m=2$, spiral modes in a galactic disk, treated as a collisionless system, by carrying out a global modal analysis within the WKB approximation. First, we study a superthin, LSB galaxy UGC~7321 and show that it does  not support discrete global spiral modes when modeled as a disk-alone system or as a disk plus dark matter system. Even a moderate increase in the stellar central surface density does not yield any global spiral modes. This naturally explains the observed lack of strong large-scale spiral structure in LSBs. An earlier work \citep{GSJ16} where the galactic disk was treated as a fluid system for simplicity had shown that the dominant halo could not arrest global   modes. We found that this difference arises due to the different dispersion relation used in the two cases and which plays a crucial role in the search for global spiral modes. 
Thus the correct treatment of stars as a collisionless system as done here results in the suppression of global spiral modes, in agreement with the observations. We performed a similar modal analysis for the Galaxy, and found that the dark matter halo has a negligible effect on large-scale spiral structure.

\end{abstract}

\end{frontmatter}
%

\section{Introduction}  
Low Surface Brightness (LSB) galaxies are characterized by low star formation rate \citep{IB97} and low disk surface density 
\citep{dBM96,dBM01}. 
The spiral structure in LSBs is often incipient or fragmentary and usually faint and difficult to trace \citep{Sch90,Mcg95,Sch11}, and they generally do not host any strong large-scale spiral structure, the kind we see in case of normal HSB galaxies like our Milky way. We note that, we are interested only in small LSBs, which are more abundant, and do not include the giant LSBs like Malin~1. Some of the giant LSBs show fairly strong, large-scale spiral structure as in UGC~6614 \citep{Das13} and in the inner regions they are dynamically similar to their High Surface Brightness (HSB) counterparts \citep{Lelli10}.  \citet{Fu02} has applied the technique of density-wave theory to put constraint on the decomposition of the rotation curve in LSBs. However we note that in the sample considered by \citet{Fu02} contains giant LSB (e.g., UGC~6614) which often show large-scale spiral structure \citep[e.g., see][]{Das13}.

 The LSBs are dark matter dominated from the very inner regions \citep{Bot97,dBM97,dBM01}. Within the optical disk, the dark matter constitutes about $90$ per cent of the total mass of LSBs, whereas for the HSBs the contributions of the dark matter halo mass and stellar mass are comparable \citep{dBM01,Jog12}. Thus, the LSBs constitute a natural laboratory to study the effect of dark matter halo on different aspects of galactic dynamics.

Several past studies have shown the effect of dominant dark matter halo in the suppression of global non-axisymmetric bar modes \citep{Mih97}, in making the galactic disks superthin \citep{BJ13} and in prohibiting the swing amplification mechanism from operating, thus explaining the lack of small-scale spiral structure as noted observationally \citep{GJ14}. 

According to the density wave theory, the grand-design spiral arms are the high density regions of a rigidly rotating spiral density wave, with a well defined pattern speed, that are self-consistently generated by the combined gravity of the unperturbed disk and the density wave \citep{LS64,LS66}. For a recent review on this see \citet{DOBA14}.

 In a recent work \citet{GSJ16} (hereafter Paper~1) investigated the role of a dominant dark matter halo on the global spiral modes within the framework of the density wave theory by treating the  galactic disk as a fluid. Using the input parameters of a superthin LSB galaxy UGC~7321 and the Galaxy, they found that for UGC~7321, the dark matter halo has a negligible effect on arresting the global spiral modes when the disk is modeled as a fluid. This is in contrast to the results for small-scale spiral features where the dark matter was shown to suppress the small-scale, swing-amplified spiral structures almost completely \citep{GJ14}. Ghosh et al. (Paper 1) argued that that since  LSBs are relatively isolated, tidal interactions are less likely to occur compared to those for the high surface brightness (HSB) galaxies.  Thus even though the global spiral modes are formally permitted in the fluid disk plus dark matter halo model, it was argued that it is the lack of tidal interaction that makes it difficult for the global spiral structure to develop in these galaxies.

In this paper we address the effect of dark matter halo on global $m=2$ modes by modeling the galactic disk more realistically as a $\emph{collisionless}$ system. A tidal encounter is likely to give rise to global modes, as has been seen in simulations, as in M51 \citep[see e.g.,][]{TT72}.
 
 We use the dispersion relation for a $\emph{collisionless}$ disk to construct global standing-wave like solutions by invoking the Bohr-Sommerfeld quantization condition (for details see Paper~1). Note that fluid disks allow wavelike solutions at small wavelengths, since fluid pressure provides the restoring force; but collisionless disks suppress wavelike modes for very small wavelengths \citep[e.g. see][]{BT87}.

The \S~2 contains formulation of the problem and the input parameters. In \S~3 we present the WKB analysis and the relevant quantization rule. \S~4 {\bf{\S~5}} contain the results and discussion, respectively while \S~6 contains the  conclusions.

\section{Formulation of the problem}
We model the galactic disk as a collisionless system characterized by an exponential surface density $\Sigma_{s}$, and one-dimensional velocity dispersion $\sigma_{\rm s}$. For simplicity, the galactic disk is taken to be infinitesimally thin. In other words, we are interested in perturbations that are confined to the mid-plane ($z=0$). The dark matter halo is assumed to be non-responsive to the gravitational perturbations of the disk. We have used cylindrical coordinates $(R, \phi, z)$ in our analysis.
\subsection{Details of models}
In this subsection, we describe the models that we have used for the study of effect of dark matter halo on the global spiral modes.

The dynamics of the disk is calculated first only under the gravity of the disk (referred to as disk-alone case) and then under the joint gravity of disk and the dark matter halo (referred to as disk plus halo case).
We took an exponential stellar disk with central surface density $\Sigma_0$  and the disk scalelength $R_{\rm d}$, which is embedded in a concentric dark matter halo whose density follows a pseudo-isothermal profile characterized by core density $\rho_0$ and core radius $R_{\rm c}$.

The net angular frequency, $\Omega$ and the net epicyclic frequency, $\kappa$ for a galactic disk embedded in a dark matter halo, concentric to the galactic disk, are given as:\\
\begin{equation}
\kappa^2= \kappa^2_{\rm disk}+\kappa^2_{\rm DM}\,; \quad \Omega^2= \Omega^2_{\rm disk}+\Omega^2_{\rm DM}\,.
\end{equation}

The expressions for $\kappa^2_{\rm disk}$, $\Omega^2_{\rm disk}$ in the mid-plane ($z=0$) for an exponential disk and $\kappa^2_{\rm DM}$, $\Omega^2_{\rm DM}$ for a pseudo-isothermal halo in the mid-plane ($z=0$) have been calculated earlier (see Paper 1 for details).

In the early-type galaxies, the bulge component dominates in the inner regions, and hence exclusion of the bulge component from the models for early-type galaxies will underestimate the rotation curve in the inner regions. In Paper 1, it was shown that for the Galaxy, inclusion of bulge yielded more realistic results (for details see \S~5.1 in Paper 1).
Also note that UGC~7321 has no discernible bulge \citep{MAT99,MAT03}. Therefore, only for our Galaxy we have included bulge in both the disk-alone and disk plus dark matter halo models. We adopt a Plummer-Kuzmin bulge model for our Galaxy which is characterized by total bulge mass $M_{\rm b}$ and bulge scalelength $R_{\rm b}$.

The expressions for $\kappa^2_{\rm bulge}$, $\Omega^2_{\rm bulge}$ in the mid-plane ($z=0$) for such a bulge having a Plummer-Kuzmin profile are given in Paper 1. Therefore, for the Galaxy, these $\kappa^2_{\rm bulge}$ and $\Omega^2_{\rm bulge}$ terms will be added in quadrature to the corresponding terms due to disk and dark matter halo.

\subsection{Input parameters}
The input parameters for different components of UGC~7321 \citep{Ban10} and the Galaxy \citep{Mer98,BLUM95} are summarized in Table 1.

\begin{table*}
\centering
\caption{Input parameters for UGC~7321 and the Galaxy}
\begin{tabular}{ccccccc}
\hline
Galaxy & $\Sigma_0$   &  $R_{\rm d}$  & $\rho_0$  & $R_{\rm c}$  & $M_{\rm b}$  & $R_{\rm b}$  \\
& (M$_{\odot}$ pc$^{-2}$) & (kpc) & (M$_{\odot}$ pc$^{-3}$) & (kpc) &  ($\times$ 10$^{10}$~M$_{\odot}$) & (kpc)\\
\hline
UGC~7321  & 50.2 & 2.1 & 0.057  & 2.5  & - & -\\
The Galaxy & 640.9 & 3.2 & 0.035 & 5 & 2.8 & 2.5\\
\hline
\end{tabular}
\end{table*} 

For UGC~7321, the stellar velocity dispersion in the radial direction is taken to be: $\sigma_{\rm s}$ =$(\sigma_{\rm s0})_R \exp(-R/2R_d)$, where $(\sigma_{\rm s0})_R$ is the central velocity dispersion in the radial direction. The observed central velocity dispersion in the $z$ direction ($(\sigma_{\rm s0})_z$) is 14.3 km sec$^{-1}$ \citep{Ban10}. In the solar neighborhood, it is observationally found that $(\sigma_{\rm s})_z$/$(\sigma_{\rm s})_R$ $\sim$ 0.5 \citep[e.g.,][]{BT87}. Here we assume the same conversion factor for all radii.

For the Galaxy, the observed stellar velocity dispersion in the radial direction is : $\sigma_{\rm s}$ =$(\sigma_{\rm s0})_R \exp(-R/8.7)$, where $(\sigma_{\rm s0})_R$ = 95 km sec$^{-1}$ \citep{LF89}.

\section{WKB analysis}
The dispersion relation for an infinitesimally thin galactic disk, modeled as a collisionless system, in the WKB limit, is given by \citep{BT87}:\\
\begin{equation}
(\omega - m\Omega)^2=\kappa^2-2\pi G \Sigma_s |k|{\mathcal F}(s, \chi),
\label{disp-equation}
\end{equation}
where  $s(=({\omega-m\Omega})/{\kappa})$  and  $\chi (=k^2\sigma^2_{\rm s}/\kappa^2)$ are the dimensionless frequencies. ${\mathcal F}(s, \chi)$ is the reduction factor which physically takes into account the reduction in self-gravity due to the velocity dispersion of stars. 
The form for ${\mathcal F}(s, \chi)$ for a razor-thin disk whose stellar equilibrium state can be described by the Schwarzchild distribution function is given by \citep{BT87}: 
\begin{equation}
{\mathcal F}(s, \chi)=\frac{2}{\chi}\exp(-\chi)(1-s^2)\sum_{n=1}^\infty\frac{I_n(\chi)}{1- s^2/n^2}\,.
\label{reduc-factor}
\end{equation}
Since we are interested in $m=2$ grand spiral modes, a rearrangement of equation (\ref{disp-equation}) yields\\
\begin{equation}
4(\Omega_{\rm p}-\Omega)^2=\kappa^2-2\pi G \Sigma_s|k| {\mathcal F}(s, \chi)\,,
\label{disp-mod}
\end{equation}
where $\Omega_{\rm p} $ (= $\omega /m$) is the pattern speed with which the spiral arm  rotates rigidly in the galactic disk.

For a given $\Omega_{\rm p}$, equation~(\ref{disp-mod}) is an implicit relation between the phase space variables $k$ and $R$. A contour of constant $\Omega_{\rm p}$ denotes the path of a wavepacket that propagates with a group velocity $v_g$ = $d\omega/dk$ \citep{Tom69}.

The simpler analytic dispersion relation for the fluid disk allows one to express $k$ explicitly as a function of $R$ (for details see \S~3 of Paper 1), but for a collisionless system, the dispersion relation (equation \ref {disp-equation}) is transcendental in nature, therefore it is impossible to analytically express the wavevector $k$ as an explicit function of $R$. However, it is straightforward to obtain this relation numerically.

\begin{figure}
\centering
\includegraphics[height=2.4in,width=3.4in]{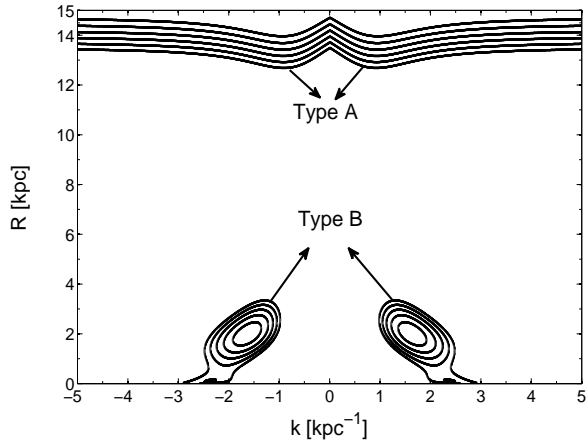}
\caption{Propagation diagrams (contours of constant $\Omega_{\rm p}$) for different pattern speeds. The input parameters used are for the Galaxy, where the Galaxy is modeled as a collisionless disk plus dark matter halo. Different types of contours present here are marked as A and B. The range of $\Omega_{\rm p}$ varies from 25.8 km s$^{-1}$ kpc$^{-1}$ to 28.7 km s$^{-1}$ kpc$^{-1}$, with a spacing of 0.5 km s$^{-1}$ kpc$^{-1}$. $\Omega_{\rm p}$ value increases from the outer to the inner contours patterns.}
\label{fig:UGC7321}
\end{figure}
Fig.~1 shows the typical contours that are present in different models considered in this work. We refer the reader to Paper 1 for a physical interpretation of these contours in terms of propagation of wavepackets.
In Paper 1, another type of contour (called {\it {Type C}}) was also present, but such contours occurred due to the fluid treatment of the disk, and consequently we did not consider them in our analysis (for details see \S~3.1 of Paper 1). As expected, these contours are absent when the disk is treated as collisionless, thus justifying our omitting them in our previous analysis.
Since global spiral modes are basically standing waves in a differentially rotating galactic disk \citep[e.g. see][]{Ber00}, therefore, in order to obtain a standing wave, the wave has to be reflected/refracted back into the wave cycle by the reflecting/refracting barrier. Although contours of {\it {Type A}} allow wavelike solutions, the wavevector $k$ can become quite large. Such waves will be dissipated at $k$ corresponding to the epicyclic length scale, and thus do not constitute valid standing waves. Hence, only a closed contour has the correct behaviour to represent a standing wave in a galactic disk. Hence, from now on, we will consider contours only of {\it {Type B}} and discard the contours of {\it {Type A}}.

The global spiral modes can be constructed from the WKB dispersion relation by using the Bohr-Sommerfeld quantization condition (for details see \S~3.1 in Paper 1).

The appropriate WKB quantization rule for the closed contours of {\it Type~B} is given by \citep{Tre01}
\begin{equation}
2 \int_{R_-}^{R_+} [k_{+}(R)-k_{-}(R)] dR = 2\pi \left (n-\frac{1}{2} \right)\,
\label{qcond}
\end{equation}
where $n=1,2,3,\cdots$; and 
 $k_{+}$, $k_{-}$ (where $k_{-} \le k_{+}$ ) are the two solutions of equation~(\ref{disp-mod}), and the equality sign holds only at the turning points ($R_{\pm}$).

\section {Results}
In principle, the pattern speed $\Omega_{\rm p}$ could be either positive or negative, implying prograde or retrograde motion of the spiral pattern,  respectively. For a negative pattern speed we can write $\Omega_p = -|\Omega_p|$, and rewrite equation (\ref{disp-mod}) in the form 
\begin{equation}
4(|\Omega_{\rm p}|+\Omega-\kappa/2)(|\Omega_{\rm p}|+\Omega+\kappa/2)=-2\pi G \Sigma_s|k|{\mathcal F}(s, \chi)\
\label {disp-negative}
\end{equation}
For both the galaxies considered here, the quantities $\Omega-\kappa/2$ and $\Omega+\kappa/2$ are positive, hence the l.h.s. of equation (\ref{disp-negative}) is always positive while the right hand side equation is always negative. Consequently, there is no possible solution of equation (\ref{disp-negative}) for negative pattern speeds. This implies 
that no global spiral modes can have negative pattern speeds,  or in other words, all  discrete global spiral modes present in different models of this paper are prograde. Therefore, all the figures displayed in this paper are for prograde spiral patterns.

Before proceeding to investigate the global modes for galaxies UGC~7321 and the Galaxy, we note that \citet{GJ14,Jog14} have calculated the Toomre Q-value \citep{Too64} for these two galaxies and have shown it to be always $>$ 1 for the input parameters used in this paper. Therefore the galactic disks of the two galaxies considered by us are stable against local axisymmetric perturbations.

\subsection{UGC~7321}

Starting from the dispersion relation (equation \ref{disp-mod}), we obtained contours of {\it{Type B}}, first for the disk-alone case, and then for the disk plus dark matter halo case, .

\subsubsection{ Disk-alone} For the disk-alone case we found closed contours of {\it{Type B}}, but on applying the quantization condition (equation {\ref {qcond}}), we found no discrete global modes, i.e., the area enclosed by the closed loops is not sufficient to satisfy equation~({\ref {qcond}}) for any integral value of $n$ (see Fig.~2).
 \begin{figure}
\centering
\includegraphics[height=2.4in,width=3.4in]{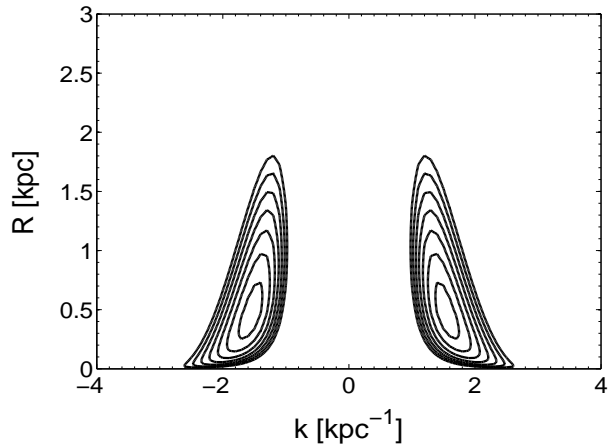}
\caption{ Propagation diagrams (contours of constant $\Omega_{\rm p}$) for the disk-alone case, corresponding to those pattern speeds which give closed loops of {\it{Type B}}. The input parameters used are for UGC~7321. The range of $\Omega_{\rm p}$ varies from 7.1 km s$^{-1}$ kpc$^{-1}$ to 7.7 km s$^{-1}$ kpc$^{-1}$, with a spacing of 0.1 km s$^{-1}$ kpc$^{-1}$. $\Omega_{\rm p}$ value increases from the outer to the inner contours patterns.}
\label{fig2}
\end{figure}

Note that the closed contours in Fig.~2 do not satisfy the condition under which WKB approximation is valid, i.e.,  $|kR| \gg 1$. In fact, some parts of the contours have $|kR| < 1$. The standard use of the WKB approximation is in quantum mechanics where it is used to compute the energy spectrum of bound systems. It is well known that at the classical turning points the WKB condition fails since $k = 0$ at the turning points. However, despite this deficiency, the approximation furnishes useful qualitative results. Note, however, that in Fig.~2 the WKB approximation does not fail so extremely. However, the failure of WKB  also implies the failure of the tight-winding approximation that makes it possible to relate the gravitational potential to the \emph{local} perturbed density. Typically, the local approximation suffices to obtain useful qualitative information even in this case, even though the numerical results are less robust. For example, in a recent paper \citet{JT12} have calculated the slow-modes of collisionless  near-Keplerian disks. They have computed eigenvalues using exact numerical methods as well as through the WKB approximation. Although their contours in some parts also do not satisfy $|kR| > 1$, they find that their results are in good qualitative agreement with the frequencies calculated using the exact method. However, note that this concern is not valid for our main result for UGC~7321 below. 

\subsubsection{Disk plus halo}
We find that for the disk plus dark matter halo case, the closed loops of {\it{Type B}} are \emph{not present} at all, and hence no discrete global modes exist in this case. This is in contrast to Paper 1 where, based on the fluid model of the disk, we found two eigenmodes for the disk plus halo case.  Since stable modes do not excite by themselves, we had argued that they would be absent in UGC 7321 due to its isolation and its consequent lack of tidal interactions with other galaxies. It is useful to note that 
it is possible to excite stable modes by tidal encounters as shown by \citet{JT12}. The difference in results can be ascribed to the difference in the treatment of the disk as a collisionless system here as opposed to a fluid  description in Paper 1. Thus, the correct treatment of stars as a collisionless system has directly led to the result that global modes are not permitted.

\subsubsection{Other LSB galaxies}
To investigate the role of dark halo on the existence of global modes in LSBs in general, we constructed models where we set the central surface density to  two and three times the value used in the original model for UGC~2371 (see Table 1), and modified the radial velocity dispersion profile in such a way that the resulting Toomre $Q$-parameter remains the same as that in the original model. The net rotation curve changes less than 20 per cent of that in the original model. Though somewhat contrived, the parameters for these additional cases were devised to check the effect of disk surface density on the existence of global modes. We did not find any global spiral modes in these modified models as well. Since a dominant dark matter halo is a characteristic feature of LSBs \citep {Bot97,dBM97}, our results indicate that low disk surface density and dark matter halo that is dominant from the innermost regions in LSBs together explain the suppression of local, swing-amplified spiral features \citep{GJ14} in these galaxies, as well as suppression of global $m=2$ spiral modes as shown here. 

In our analysis, we have used the input parameters of the LSB galaxy UGC~7321, which is an edge-on galaxy (angle of inclination 88$^{\circ}$, \citet{MA00}). However, we note that the generic features of typical LSBs that play the dominant role in the dynamics of disks in LSBs, i.e., low surface density and dark matter dominance from the innermost radius, are both observed in UGC~7321. Due to the edge on nature of this galaxy it is difficult to conclude if large scale spiral features are absent in this galaxy, although the absence of dust lanes is suggestive of this fact. In any case, for completeness of analysis we have also considered synthetic models where the disk mass was artificially enhanced by a factor of few. Therefore, the main result of this paper that the dominant dark matter halo suppresses the large-scale spiral features in LSBs is likely to hold for all LSB galaxies. 

\subsection{The Galaxy}

As a check, we carried out the same modal analysis to examine the effect of the dark matter halo on the global spiral modes for the Galaxy. The possible closed contours of {\it{Type B}} for both disk-alone and disk plus dark matter halo are shown in Fig.~3.
\begin{figure}
\centering
\includegraphics[height=2.4in,width=3.4in]{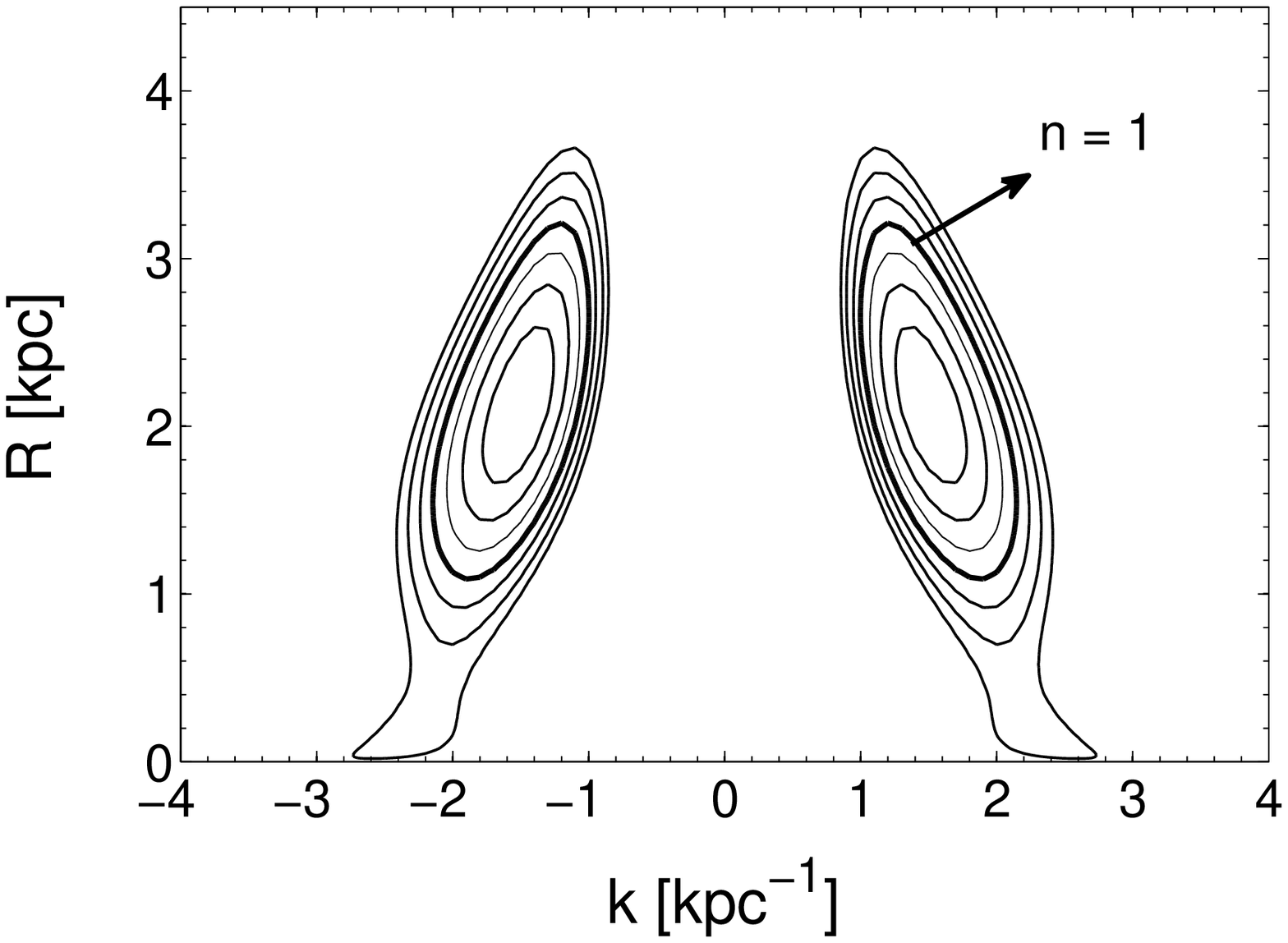}
\medskip
\includegraphics[height=2.4in,width=3.4in]{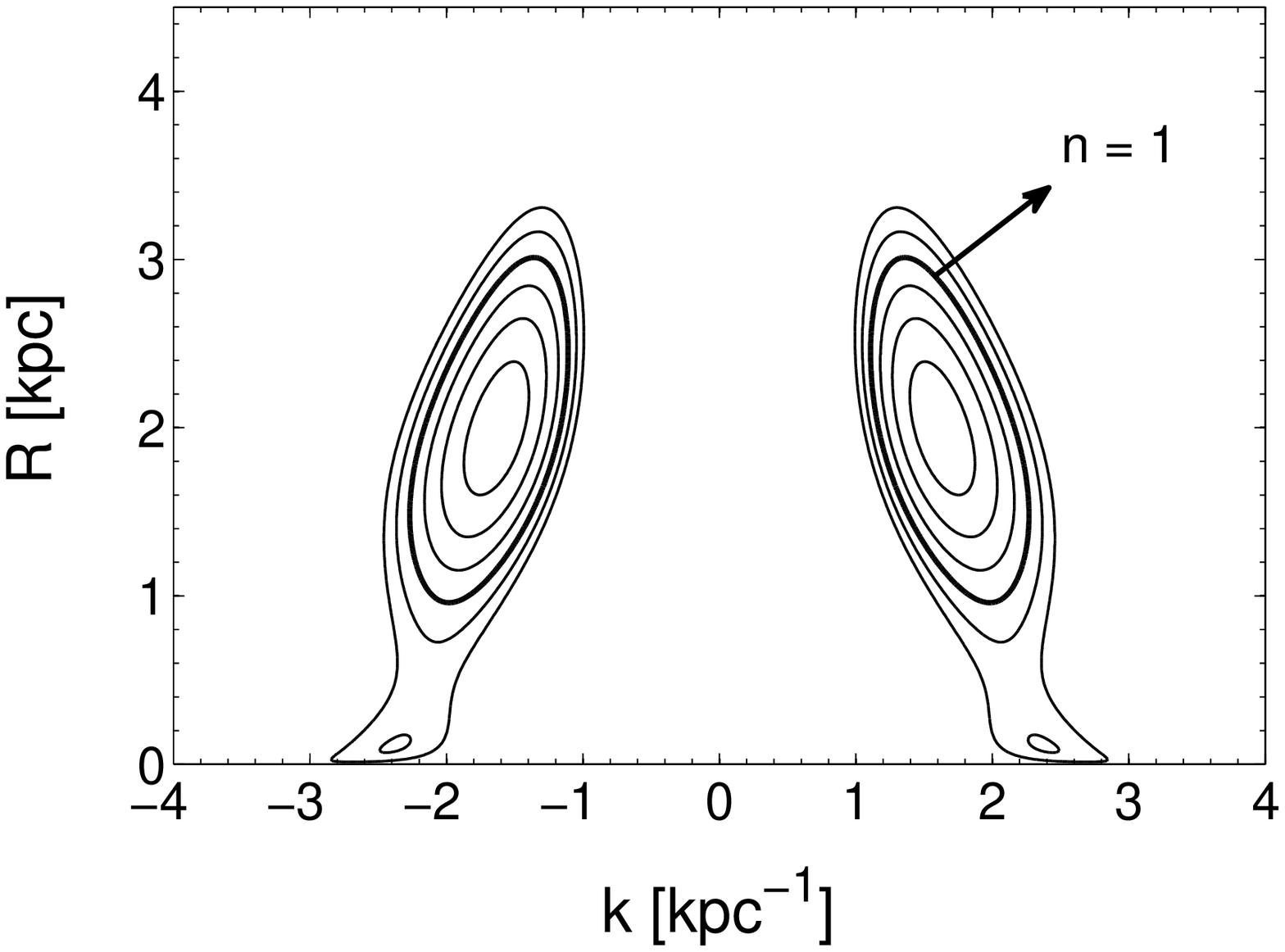}
\caption{ Propagation diagrams (contours of constant $\Omega_{\rm p}$) for those pattern speeds which give closed loops of {\it{Type B}}. The input parameters used are for The Galaxy. The top panel shows contours for disk-alone case (including the bulge) where the range of $\Omega_{\rm p}$ varies from 26.1 km s$^{-1}$ kpc$^{-1}$ to 28.8 km s$^{-1}$ kpc$^{-1}$, at intervals of 0.4 and the bottom panel shows the contours for disk plus dark matter halo case (including the bulge) where the range of $\Omega_{\rm p}$ varies from 25.9 km s$^{-1}$ kpc$^{-1}$ to 28.1 km s$^{-1}$ kpc$^{-1}$, at intervals of 0.4. The closed loops that correspond to the global modes for different models, are indicated by solid lines. $\Omega_{\rm p}$ value increases from the outer to the inner contours.}
\label{fig3}
\end{figure}
Next, we applied the quantization condition to seek the discrete global mode(s), and the results are summarized in Table~2. Also the positions of different resonance points for the two modes in Table~2 are presented Fig.~\ref{fig4}.\\
\begin{table*}
\centering
 \begin{minipage}{.95 \textwidth}
\caption{Results for global modes for the Galaxy}
\begin{tabular}{cccccc}
\hline
Case& $\Omega_{\rm p}$  &  $R_- $  & $R_+$ & $R_{\rm CR}$ & $n$ \\
&(km s $^{-1}$ kpc$^{-1}$) &  (kpc) & (kpc) & (kpc) & \\
\hline
Disk-alone case (including bulge)& 27.3 & 1.1  & 3.3 & 8.3 & 1\\
Disk plus halo case (including bulge) & 26.7  & 0.9 & 3.0 & 8.6 & 1 \\
\hline
\end{tabular}
\end{minipage}
\end{table*}

\begin{figure}
\centering
\includegraphics[height=2.4in,width=3.4in]{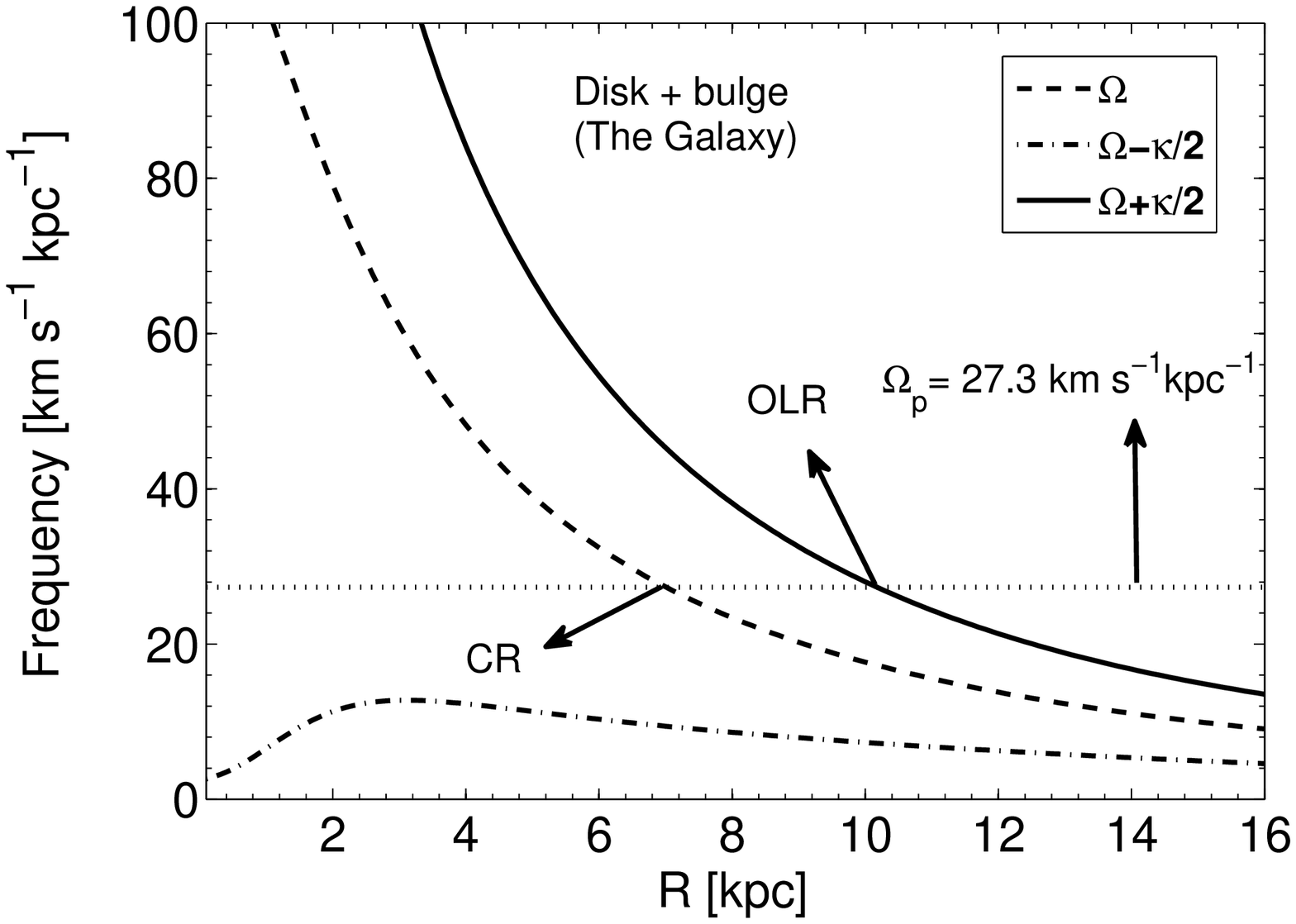}
\medskip
\includegraphics[height=2.4in,width=3.5in]{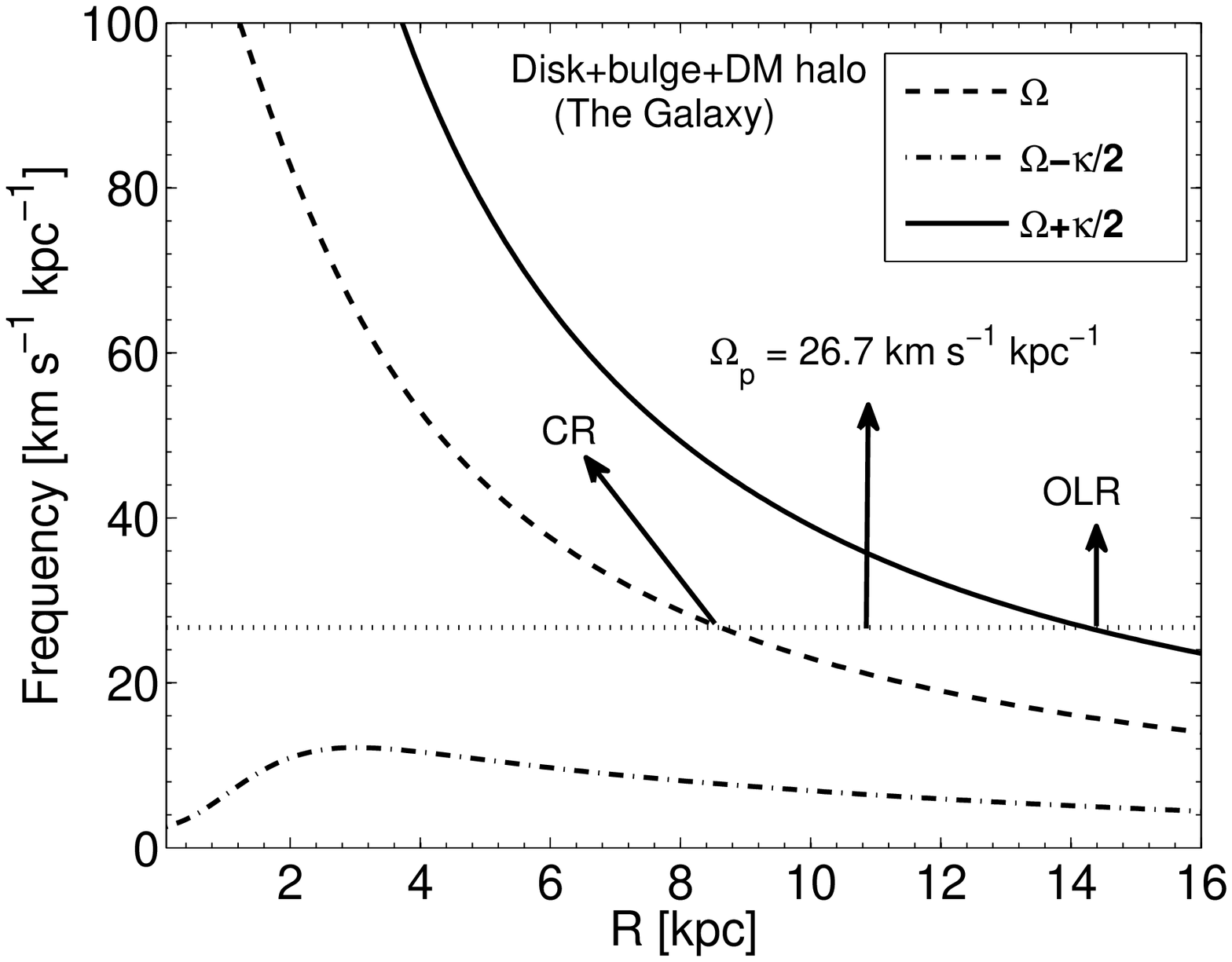}
\caption{Positions of different resonance points for the two modes in Table~2 are presented here. Comparison with Fig.~\ref{fig3} shows that the modes do not extend up to the resonance points, consistent with the fact that $k$ is not equal to zero anywhere for modes in Fig.~\ref{fig3}.}
\label{fig4}
\end{figure}

From Table~2, it is evident that the dark matter halo has a negligible effect on the global spiral modes for the Galaxy, the difference can be seen only in the change of the specific values of pattern speed $\Omega_{\rm p}$. This result is at par with our expectation, since the dark matter halo is known to be not important in the inner regions of the Galaxy \citep [e.g., ][]{Sac97,dBM01}. These results are consistent with the results of Paper 1 where the Galactic disk was modeled as a fluid disk.

The key difference with results from  Paper 1 is that the specific values of pattern speed $\Omega_{\rm p}$ which give the global modes are different   
(for comparison, see Table 3 in Paper 1). \citet {Sie12} found a range of $18-24$ km s$^{-1}$ kpc$^{-1}$ for the pattern speed of spiral arms for the Galaxy. On comparison the pattern speed values obtained in this paper lie outside the observed range of pattern speed. 

The shape of the closed contours in this work (Figs. 1-3) is different from that in Paper 1. A curve of constant $\Omega_{\rm p}$ had nearly identical $k$ values at the turning points of a closed curve of {\it {Type B}} in Paper 1, whereas the contours look slanted in the present paper, meaning that the $k$ values at the turning points are substantially different. This is due to the difference in the dispersion relation for the
two cases. A more detailed interpretation is not possible in a simple analytical form, since the dispersion relation for the collisionless case (equation~(\ref{disp-equation})) has a transcendental form.

In this paper for simplicity we have not included the low velocity dispersion component, namely, gas in the system which could further modify the obtained range of pattern speeds. Note that any late-type spiral galaxy contains non-negligible amount of gas, and it has been shown that the gas plays a  significant role in various dynamical issues \citep[e.g.,][]{JS84a,JS84b,Jog92,Raf01,GJ15,GJ16}. The amount of gas present in UGC~7321 is quite small in comparison to that in HSB galaxies \citep{UM03}. Therefore, the results for UGC~7321 obtained in this paper - in particular the suppression of global modes by the dominant dark matter halo is unlikely to change significantly if gas is included in the calculation. The effect of gas on the global spiral modes in a gravitationally coupled two-component (stars and gas) system, as applicable to the gas-rich HSB galaxies, will be taken up in a future paper.

\section{Discussion}
The results presented here are based on the assumption that the grand-design spiral structure seen in disk galaxies is due to the density waves, as proposed by \citet{LS64,LS66}. However this hypothesis has not yet been fully confirmed observationally. For example, the angular off-set in age of the stellar population, as predicted by the classical density wave theory has not been found in studies by \citet{Foy11,Fer12} in their sample of late-type disk galaxies. Also the study of $CO$ observation of NGC~1068 using generalized Tremaine-Weinberg method by \citet{MRM06} has revealed that the pattern speed of the spiral structure of this galaxy varies rapidly with radius, indicative of short-lived spiral features. A study of the gas content and the $H\alpha$ line of sight velocity distribution in NGC~6754 by \citet{San16} revealed a different sense of streaming motion in the trailing and leading edge of this galaxy, indicating that the spiral arms of this galaxy is likely to be a transient. 

On the other hand, it is also worth noting that \citet{Im16} have recently presented a study of measuring pitch angle from the images taken in different wavelengths for a large sample of disk galaxies and found that the pitch angle varies from one wavelength to another as predicted by the classical density wave theory, and thus furnishes strong observational evidence for the validity of the density-wave theory of spiral structure in disk galaxies.

Generally the various $N$-body studies so far do not show the evidence of long-lived spiral structure. Several $N$-body simulations have shown that the spiral arms are transient and get wound up quickly \citep{Sel11,Fuji11,Gra12,Don13}, which goes against the classical density wave picture. Interestingly an opposite trend is shown in a recent work by \citet{SaEl16} who reported long-lived spiral structure in a $N$-body study of models for galaxies with intermediate bulges.

\section {Conclusion}
We have investigated the effect of a dominant dark matter halo on the possible existence of global spiral arms while modeling the galactic disk as a \emph {collisionless} system. The WKB dispersion relation and  the Bohr-sommerfeld quantization condition were used to obtain discrete global spiral modes present in any model. We have analysed a superthin LSB galaxy UGC~7321, and the Galaxy, for this work. We found that for UGC~7321 both the disk-alone and the disk plus dark matter halo cases did not yield any discrete global spiral modes. Even an increase in the stellar central surface density by a factor of few failed to produce any global spiral modes. Thus our findings provide a natural explanation for the observed dearth of  strong large-scale spiral structure in the LSBs. Our results differ from those obtained in Paper 1 where the galactic disk was treated as a fluid, where it was found that the dominant dark matter halo had a  negligible 
constraining effect on the existence of global spiral modes in these galaxies. This difference is due to the different dispersion relation for fluid and collisionless systems, which played a pivotal role in determining the discrete global mode(s) present in a model. 
Thus the correct collisionless treatment for stars as done here has led us to a result for the non-existence of global spiral modes which
is in agreement with the observations.  
As a check, for the Galaxy we carried out  a similar modal analysis and found that the dark matter halo has a negligible effect on the global spiral modes, as expected since the Galaxy is not dark matter dominated in the inner regions, in contrast to UGC~7321. 

Thus, the dark matter halo that dominates from the innermost regions is shown to suppress the growth of local, swing-amplified non-axisymmetric features \citep{GJ14} as well as the global spiral modes as shown here.

\bigskip
\noindent {\bf Acknowledgements:} 
We thank the anonymous referee for the constructive comments that have helped to improve the paper.
CJ would like to thank the DST, Government of India for support via 
J.C. Bose fellowship (SB/S2/JCB-31/2014).

\end{document}